\documentclass[iicol,sn-mathphys]{sn-jnl}
\usepackage{graphicx}

\newcommand{\beq}{\begin{equation}}
\newcommand{\eeq}{\end{equation}}
\newcommand{\beqa}{\begin{eqnarray}}
\newcommand{\eeqa}{\end{eqnarray}}

\newcommand{\cluster}{{\rm cl}}
\newcommand{\dd}{{\rm d}}
\renewcommand{\e}{{\rm e}}
\renewcommand{\max}{{\rm max}}
\newcommand{\mean}[1]{\left\langle#1\right\rangle}
\newcommand{\mic}{{\rm mic}}
\renewcommand{\middle}{{\rm mid}}
\renewcommand{\min}{{\rm min}}
\newcommand{\open}{{\rm open}}
\newcommand{\prob}{{\rm Prob}}
\newcommand{\s}{\sigma}
\newcommand{\tr}{\mathop{\rm tr}}
\renewcommand{\u}{{\bullet}}
\renewcommand{\z}{{\circ}}

\renewcommand{\C}{{\cal C}}
\renewcommand{\H}{{\cal H}}
\newcommand{\N}{{\cal N}}
\newcommand{\E}{{\bf E}}
\renewcommand{\L}{{\bf L}}
\renewcommand{\R}{{\bf R}}
\renewcommand{\T}{{\bf T}}
\newcommand{\Y}{{\bf Y}}
\renewcommand{\Z}{{\cal Z}}

\begin{document}

\title
[From the Ising chain to the Riviera model]
{Jamming and metastability in one dimension:
from the kinetically constrained Ising chain to the Riviera model}

\author[1]{\fnm{P.~L.} \sur{Krapivsky}}\email{pkrapivsky@gmail.com}
\author*[2]{\fnm{J.~M.} \sur{Luck}}\email{jean-marc.luck@ipht.fr}

\affil[1]{\orgdiv{Department of Physics}, \orgname{Boston University},
\city{Boston}, \postcode{MA 02215}, \country{USA}}
\affil[2]{\orgdiv{Institut de Physique Th\'eorique},
\orgname{Universit\'e Paris-Saclay, CNRS \& CEA},
\city{91191~Gif-sur-Yvette}, \country{France}}

%\date{\today}

\abstract{
The Ising chain with kinetic constraints
provides many examples of totally irreversible zero-temperature dynamics
leading to metastability with an exponentially large number of attractors.
In most cases, the constrained zero-temperature dynamics
can be mapped onto a model of random sequential adsorption.
We provide a brief didactic review,
based on the example of the constrained Glauber-Ising chain,
of the exact results on the dynamics of these models and on their attractors
that have been obtained by means of the above mapping.
The Riviera model introduced recently by Puljiz et~al.
behaves similarly to the kinetically constrained Ising chains.
This totally irreversible deposition model however does not enjoy the shielding property
characterising models of random sequential adsorption.
It can therefore neither be mapped onto such a model
nor (in all likelihood) be solved by analytical means.
We present a range of novel results on the attractors of the Riviera model,
obtained by means of an exhaustive enumeration for smaller systems
and of extensive simulations for larger ones,
and put these results in perspective with the exact ones
which are available for kinetically constrained Ising chains.}

\maketitle

\section{Introduction}
\label{intro}

The slow non-equilibrium dynamics of a broad variety of systems,
including structural glasses, spin glasses and granular media,
has often been modelled in terms of the motion of a particle
in a multi-dimensional energy landscape with a great many valleys
separated by barriers~\cite{gold}.
In mean-field models, barrier heights diverge in the thermodynamic limit,
and so valleys become truly metastable states~\cite{thou,kirk,mpv}.
For finite-dimensional systems with short-range interactions,
barrier heights and valley lifetimes remain finite at any finite temperature,
so that metastability becomes a matter of time scales~\cite{gotze,gs,bm,debe,bb,bg}.

Zero-temperature dynamics may also lead to metastability,
in the broad sense that the system does not reach its ground state.
Focussing for definiteness our attention onto spin models on finite-dimensional lattices,
zero-temperature dynamics
is known to lead to metastability in several instances.
The two-dimensional ferromagnetic Ising model under zero-temperature Glauber dynamics
may end in configurations with one or more frozen-in stripes~\cite{skr1,skr2},
whose statistics has been described via crossing probabilities
in critical percolation~\cite{bkr,okr}.
On higher-dimensional lattices, the pattern of possible attractors is richer,
including configurations with stochastic blinkers,
i.e., clusters of spins which flip forever~\cite{skr1,okr3}.
The ferromagnetic spin chain under Glauber dynamics~\cite{glau}
reaches its ground state according to the coarsening paradigm~\cite{bray},
whereas zero-temperature Kawasaki dynamics yields metastability,
in the sense that the energy of the blocked (or jammed) configurations
reached by the dynamics is extensively above the ground-state energy~\cite{cks,dsk}.
Spin chains endowed with zero-temperature single-spin-flip dynamics
may also manifest metastability, at least in the following in two instances.
In the presence of random ferromagnetic couplings,
the spin chain has an exponentially large number
of metastable configurations~\cite{dg,msj}.
On the other hand, preventing energy-preserving spin flips,
and more generally imposing kinetic constraints,
often leads to metastability.
The latter situation is considered in detail below.

Kinetically constrained models have been introduced as the simplest systems of all
exhibiting aging and many other features of the phenomenology of glassy dynamics~\cite{rs}.
For a broad range of one-dimensional kinetically constrained Ising models,
zero-temperature dynamics exhibits metastability in the strong sense,
with an exponentially large number of attractors,
which are blocked (or jammed) configurations~\cite{pfa,pje,pse,pf,ef,kpri,klin,kkra,dl1,dl2,dl3,pb,ds,gl}.
These attractors often have a simple local characterization
in terms of forbidden patterns.
As a consequence, the configurational entropy~$S(E)$~\cite{jackle,palmer},
such that the number of attractors at fixed energy density~$E$ grows as
\beq
\N(E)\sim\exp(NS(E)),
\label{entdef}
\eeq
can be determined exactly,
either by a direct combinatorial approach,
by the transfer-matrix method used hereafter,
or by a more recent renewal method~\cite{renewal},
depending on the model.

Extended stochastic dynamical systems
such as the above kinetically constrained models,
having an exponentially large number of attractors,
are usually strongly non ergodic.
In particular, the energy density $E_\infty$ at the end of the dynamics,
where the system has settled into one of the many attractors which are at its disposal,
depends continuously on the energy density $E_0$ of the initial state.
One is then entitled to ask more detailed questions
concerning the dynamical weights of attractors.
If the system is launched from a random initial configuration with energy density $E_0$,
does it sample all attractors with energy density~$E_\infty$
with equal weights, i.e., with a flat measure,
or, on the contrary, do the shape and size of the attraction basin
of every single attractor matter?
This question has been first addressed for slowly compacting granular matter.
In this context, the assumption that the system explores its phase space
with a flat measure is referred to as the Edwards hypothesis~\cite{edbook,edm,med,edo}
(see~\cite{edrev} for a recent review).

The one-dimensional kinetically constrained Ising models
investigated in~\cite{pfa,pje,pse,pf,ef,kpri,klin,kkra,dl1,dl2,dl3,pb,ds,gl}
have the remarkable property that their zero-temperature dynamics
can be mapped onto a model of random sequential adsorption (RSA)
or cooperative sequential adsorption (CSA)~\cite{evans,talbot,kinetic},
where identical objects are deposited sequentially and irreversibly on a substrate.
These models are extensions of two well-known historical examples,
the dimer deposition model on the chain solved by Flory~\cite{flory}
and the parking model on the continuous line solved by R\'enyi~\cite{renyi}.
The constrained Ising chain reviewed in Section~\ref{ising}
maps onto Flory's dimer deposition model.
One-dimensional RSA and similar models enjoy a peculiar property dubbed shielding:
the deposition of any elementary object splits the line into two half-lines
whose future evolutions are independent from each other.
The latter property ensures the exact solvability of this class of models,
by means of a common analytical approach based on considering empty intervals.
Many quantities concerning the dynamics of the system
and its attractors can be derived in this way.
Let us mention the dependence of the final energy density $E_\infty$
on the initial one $E_0$,
which manifests the strong non-ergodicity of the dynamics,
and the peculiarities of connected correlation functions,
which generically exhibit oscillations and,
more importantly, a superexponential, inverse-factorial fall-off.
These results have established in particular that the
Edwards hypothesis is in general not exactly valid
for one-dimensional kinetically constrained models.

The goal of the present paper is to revisit the theory of
one-dimensional spin models with kinetic constraints from a didactic viewpoint,
and to use this body of knowledge to shed some new light
onto the Riviera model introduced recently~\cite{hr3}.
The setup of this paper is as follows.
In Section~\ref{ising} we consider in detail the prototypical example
of the kinetically constrained Ising chain,
which can be mapped onto Flory's dimer deposition model.
We successively review the various {\it a priori} statistical ensembles
of attractors (Section~\ref{isingstat})
and the exact results that can be derived on the dynamics of the model
and on the statistics of its attractors (Section~\ref{isingdy}).
Section~\ref{riv} is devoted to the Riviera model.
The latter model is a one-dimensional variant,
introduced recently by Puljiz et~al.~\cite{hr3},
of a two-dimensional irreversible deposition model
introduced earlier by the same authors~\cite{hr1,hr2}.
Houses are sequentially built on an infinite array of pre-drawn plots along a beach,
with the constraint that every house should enjoy the sunlight
from at least one of the side directions.
In other words, {\it at least one} of the two neighboring plots
of each house should remain forever unbuilt.
The strand is initially empty.
New houses are successively introduced until a blocked configuration is reached.
The above dynamical rule couples both neighboring sites of any occupied one.
The Riviera model therefore does not enjoy the shielding property.
The ensuing lack of exact solvability has observable consequences,
including the exponential decay of the connected occupation correlation.
We shall first consider
the various {\it a priori} statistical ensembles of attractors (Section~\ref{rivstat}),
and then present two kinds of results on the statistics of attractors,
namely exact results obtained by enumeration on small systems (Section~\ref{rivexact}),
and numerical results in the thermodynamic limit of very large systems (Section~\ref{rivnum}).
Section~\ref{disc} contains a discussion of our main findings.

\section{The kinetically constrained Ising chain}
\label{ising}

\subsection{Generalities}
\label{isinggal}

The reduced Hamiltonian of the ferromagnetic Ising chain reads
\beq
\H=-\sum_n\s_n\s_{n+1},
\label{ham}
\eeq
where $\s_n=\pm1$ are classical Ising spins living on the sites of the infinite chain.

To set the stage,
the Ising chain is endowed with single-spin-flip dynamics,
where each spin is flipped in continuous time at a rate $w(\delta\H)$
depending only on the energy difference caused by the flip,
\beq
\delta\H=2\s_n(\s_{n-1}+\s_{n+1})\in\{-4,0,+4\}.
\eeq
Imposing detailed balance at inverse temperature~$\beta$ gives one single relation
between the three rates, namely
\beq
w(+4)=\e^{-4\beta}w(-4).
\eeq

At zero temperature, we have $w(+4)=0$,
expressing that spin flips which would increase the energy are suppressed.
Choosing time units such that $w(-4)=1$,
zero-temperature dynamics is therefore entirely characterized by the rate $w(0)$
of energy-conserving flips.
For generic values of the latter rate,
the model exhibits coarsening~\cite{bray}.
This is in particular the case for
the historical example of Glauber dynamics~\cite{glau},
corresponding to $w(0)=1/2$.
The typical size of ferromagnetically ordered domains grows asymptotically as
$L(t)\sim\sqrt{t}$,
and so the energy density above the ground-state energy falls off as $1/\sqrt{t}$.

Energy-conserving flips are suppressed for the particular choice $w(0)=0$.
The descent dynamics thus obtained defines the kinetically constrained Ising chain.
The only possible flips are those involving isolated spins:
\beq
-+-\to---,\quad +-+\to+++.
\label{flips}
\eeq
Isolated spins therefore flip once during their whole history,
whereas other spins never flip.
This is one of the simplest kinetically constrained models,
which has been investigated at length in~\cite{dl1,dl2,dl3,pb,ds,gl}.
The main outcomes of these works will be reviewed in this section.

The descent dynamics of the constrained Ising chain can be recast as follows.
We consider the dual lattice, where original bonds are mapped onto dual sites,
and we encode satisfied (resp.~unsatisfied) bonds as occupied (resp.~empty) dual sites:
\beq
\begin{matrix}
\u:\ \tau_n=\s_n\s_{n+1}=+1,\cr
\z:\ \tau_n=\s_n\s_{n+1}=-1,
\end{matrix}
\eeq
so that the Hamiltonian~(\ref{ham}) reads
\beq
\H=-\sum_n\tau_n.
\eeq
The spin flips~(\ref{flips}) translate to
\beq
\z\z\to\u\u.
\eeq

The constrained dynamics is thus mapped onto the dimer deposition model.
The latter model, which has been first solved by Flory~\cite{flory},
is the simplest among all models in the RSA class.
Attractors of the constrained Ising chain
are in one-to-one correspondence with blocked configurations of the dimer deposition model.
The energy density~$E$ of the Ising chain is related to the particle density $\rho$
of the dimer model as
\beq
E=1-2\rho.
\eeq
We shall focus our attention
onto the two-point energy correlation function
\beq
C_n=\mean{\tau_0\tau_n}.
\label{gdef}
\eeq

\subsection{Statistical ensembles of attractors}
\label{isingstat}

In terms of the dimer deposition model,
the attractors are all the configurations where empty sites are isolated.
Equivalently, in terms of the Ising chain, there are no isolated spins.
The statistical ensemble where all attractors are equally probable,
according to a flat Edwards measure,
is referred to as the full {\it a priori} ensemble,
whereas the ensemble consisting of all attractors with fixed energy density $E$
with equal weights is the microcanonical {\it a~priori} ensemble.

\subsubsection{Transfer-matrix formalism}

As usual in statistical physics,
instead of considering the microcanonical ensemble at fixed energy density $E$,
it is advantageous to introduce the canonical ensemble
defined by introducing an effective inverse temperature~$\beta$
conjugate to the Hamiltonian $\H$.
For a finite chain of $N$ sites, we thus introduce the partition function
\beq
Z_N=\sum_\C\e^{-\beta\H(\C)},
\eeq
where the sum runs over all blocked configurations~$\C$ of a system of $N$ sites.
The partial partition functions $Z_N^\u$ and~$Z_N^\z$,
defined by assigning fixed values to the occupation of the rightmost site,
obey linear recursions of the form
\beq
\begin{pmatrix}
Z_{N+1}^\u\cr Z_{N+1}^\z
\end{pmatrix}
=\T
\begin{pmatrix}
Z_N^\u\cr Z_N^\z
\end{pmatrix},
\eeq
involving the $2\times2$ transfer matrix
\beq
\T=
\begin{pmatrix}
\e^\beta & \e^\beta \cr \e^{-\beta} & 0
\end{pmatrix}.
\eeq
The eigenvalues of $\T$ read
\beq
\lambda_\pm=\frac12\left(\e^\beta\pm\sqrt{4+\e^{2\beta}}\right).
\label{egvs}
\eeq
An associated bi-orthogonal set of left and right eigenvectors,
such that
\beq
\L_\pm\cdot\R_\pm=1,\quad\L_\pm\cdot\R_\mp=0,
\label{egor}
\eeq
reads
\beq
\L_\pm=\frac{
\begin{pmatrix}
\lambda_\pm & \e^\beta
\end{pmatrix}
}{\lambda_\pm^2+1}
,\quad
\R_\pm=
\begin{pmatrix}
\lambda_\pm \cr \e^{-\beta}
\end{pmatrix}.
\label{egfs}
\eeq

\subsubsection{Configurational entropy}

\begin{figure}[!ht]
\begin{center}
\includegraphics[angle=0,width=1\linewidth,clip=true]{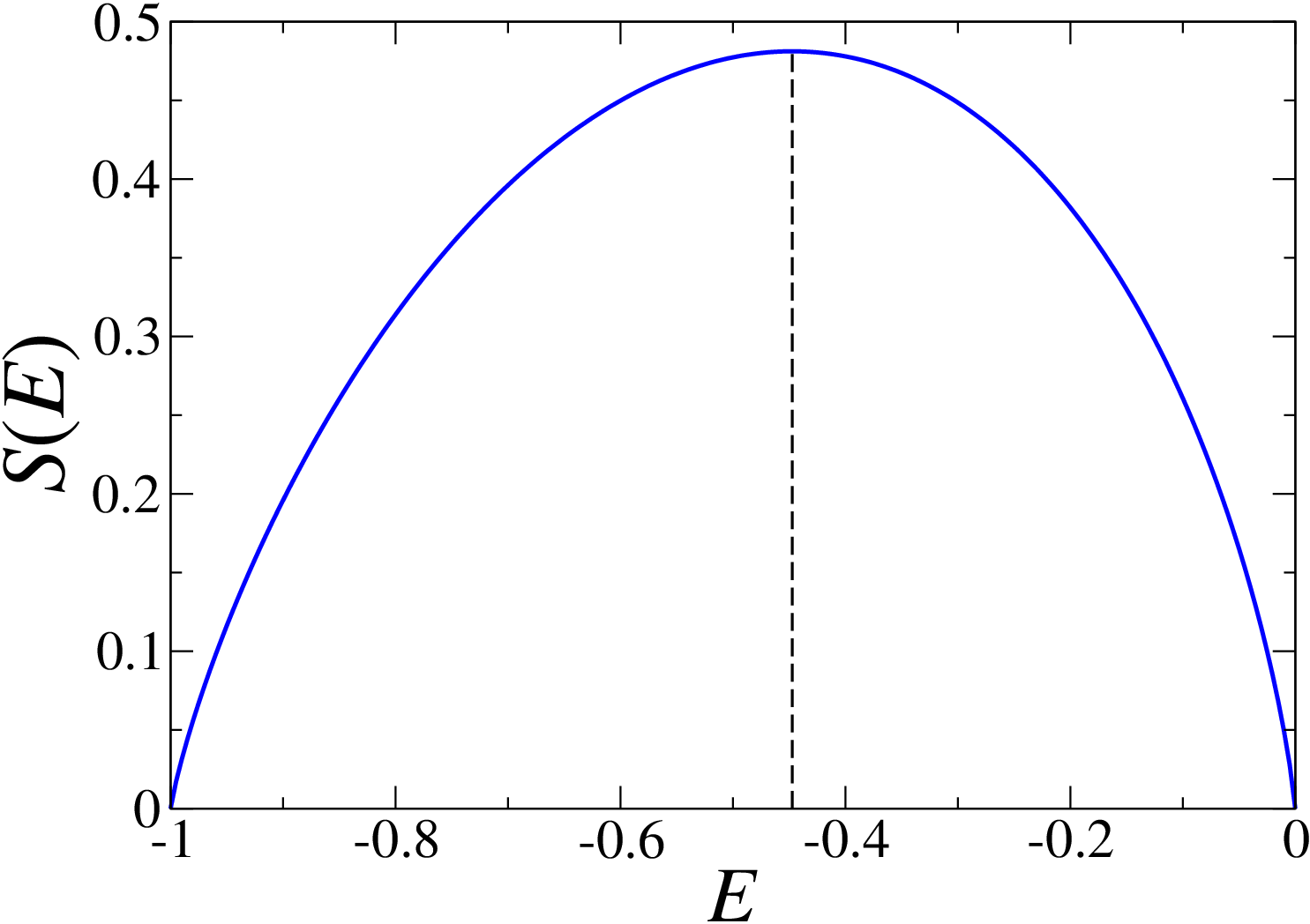}
\caption{
Configurational entropy $S(E)$ of the constrained Ising chain,
as given in~(\ref{sentres}),
against energy density $E$ of attractors (blocked configurations).
Vertical dashed line: {\it a priori} energy density $E_\star$
yielding the maximum configurational entropy $S_\star$
(see~(\ref{sap}),~(\ref{eap})).}
\label{diment}
\end{center}
\end{figure}

The configurational entropy $S(E)$ introduced in~(\ref{entdef})
can be derived from the transfer-matrix formalism as follows.
In the thermodynamic limit of a very large system size~$N$, we have
\beqa
Z_N&\approx&\int\N(E)\e^{-\beta NE}\dd E
\nonumber\\
&\sim&\int\e^{N(S(E)-\beta E)}\dd E
\nonumber\\
&\sim&\e^{N\ln\lambda_+(\beta)}.
\eeqa
We thus find that $S(E)$ is the Legendre transform of $\ln\lambda_+(\beta)$, i.e.,
\beq
S(E)=\ln\lambda_+(\beta)+\beta E,
\label{sleg}
\eeq
with
\beq
\beta=\frac{\dd S}{\dd E},\quad E=-\frac{\dd\ln\lambda_+}{\dd\beta}.
\label{leg}
\eeq
The above relations are germane
to the usual thermodynamic ones between internal energy $E$ and free energy $F=E-TS$.
Using~(\ref{egvs}), we obtain the expressions
\beq
E=-\frac{\e^\beta}{\sqrt{4+\e^{2\beta}}},
\quad
\e^\beta=-\frac{2E}{\sqrt{1-E^2}},
\label{egids}
\eeq
and
\beqa
S(E)
&=&\frac12(1-E)\ln(1-E)
\nonumber\\
&-&\frac12(1+E)\ln(1+E)
\nonumber\\
&+&E\,\ln(-2E).
\label{sentres}
\eeqa
The configurational entropy $S(E)$ is plotted
against the energy density $E$ in Figure~\ref{diment}.
It vanishes at both endpoints $E_\min=-1$ and $E_\max=0$.
Its maximum, corresponding to $\beta=0$, namely
\beq
S_\star=\ln\frac{\sqrt{5}+1}{2}\approx0.481212,
\label{sap}
\eeq
is reached for
\beqa
E_\star&=&-\frac{\sqrt{5}}{5}\approx-0.447214.
\label{eap}
\eeqa
The above number is therefore the {\it a priori} most probable value
of the energy density of an attractor of the Ising chain.

Another interpretation of the configurational entropy $S(E)$ is as follows.
The {\it a priori} probability $P(E)$ of observing
in the full statistical ensemble a blocked configuration
with an atypical energy density $E\ne E_\star$
is given by a large-deviation formula of the form
\beq
P(E)\sim\exp(-N\Sigma(E)),
\eeq
where the {\it a priori} large-deviation function reads
\beq
\Sigma(E)=S_\star-S(E).
\label{sigap}
\eeq
This quantity is positive and vanishes quadratically in the vicinity of $E=E_\star$.

\subsubsection{Correlation function}

The expression $C_{n,\mic}$ of the two-point correlation introduced in~(\ref{gdef})
in the microcanonical {\it a priori} ensemble at fixed energy density~$E$
can also be derived from the transfer-matrix formalism.
We have for $n\ge0$
\beq
C_{n,\mic}=\lim_{M,N\to\infty}\frac{\tr\T^M\E\T^n\E\T^N}{\tr\T^{M+n+N}},
\eeq
i.e.,
\beqa
C_{n,\mic}
&=&\frac{\L_+\cdot\E\T^n\E\R_+}{\lambda_+^n},
\nonumber\\
&=&(\L_+\cdot\E\R_+)^2
\label{eqgap}
\\
&+&(\L_+\cdot\E\R_-)(\L_-\cdot\E\R_+)\left(\frac{\lambda_-}{\lambda_+}\right)^n.
\nonumber
\eeqa
According to the equivalence between statistical ensembles,
the transfer matrix~$\T$, its largest eigenvalue $\lambda_+$,
and the associated eigenvectors $\L_+$ and $\R_+$
are evaluated at the effective inverse temperature $\beta$
related to the energy density~$E$ by~(\ref{egids}).
The diagonal matrix
\beq
\E=
\begin{pmatrix}
-1&0 \cr 0&1
\end{pmatrix}
\eeq
is the energy operator.

Using~(\ref{egvs}),~(\ref{egor}),~(\ref{egfs}), as well as
\beq
\L_\pm\cdot\E\R_\pm=\pm E,\quad\L_\pm\cdot\E\R_\mp=1\pm E,
\eeq
we are left with the simple result
\beq
C_{n,\mic}=E^2+(1-E^2)\left(-\frac{1+E}{1-E}\right)^n.
\label{gap}
\eeq
The disconnected part of the above correlation
is the square of the energy density $E$, as should be.
The connected part exhibits exponentially damped oscillations.

\subsection{Exact results on dynamics and attractors}
\label{isingdy}

In this section we review,
on the example of the constrained Ising chain
whose dynamics can be mapped onto the dimer deposition model~\cite{dl1,dl2,dl3,pb,ds,gl},
the approach yielding exact analytical results
on the dynamics and the attractors of RSA models in one dimension
(see~\cite{evans,talbot,kinetic} for reviews).

We consider a factorized initial state where each variable $\tau_n$ is chosen at random as
\beq
\begin{matrix}
\tau_n=-1 & (\z)\quad & \hbox{with prob.} & p,\hfill\cr
\tau_n=+1 & (\u)\hfill & \hbox{with prob.} & 1-p,\cr
\end{matrix}
\eeq
so that the energy density of the Ising chain
and the particle density of the dimer model read
\beq
E(0)=E_0=2p-1,\quad\rho(0)=\rho_0=1-p.
\label{erzero}
\eeq
In particular, $p=1$ corresponds to the dimer system being initially empty,
whereas $p=1/2$ corresponds to an infinite-temperature initial state for the Ising chain.

\subsubsection{Energy density}
\label{isingdye}

The temporal evolution of thermodynamic quantities such as the energy density $E(t)$
can be derived by analytical means
by using the shielding property of one-dimensional RSA and germane problems,
implying that the densities of empty intervals obey closed evolution equations.
In terms of the dimer deposition model, let
\beq
p_\ell(t)
=\prob(\u\underbrace{\z\cdots\z}_{\ell}\u)
\label{pt}
\eeq
denote the density per unit length at time $t$ of intervals
consisting of exactly $\ell$ consecutive empty sites, with $\ell\ge1$,
so that we have
\beq
E(t)=1-2\rho(t),\quad
\rho(t)=1-\sum_{\ell\ge1}\ell p_\ell(t).
\eeq
The probabilities $p_\ell(t)$
obey evolution equations which can be derived as follows.
Clusters of length $\ell=1$ are inactive,
whereas a new dimer can be deposited at $(\ell-1)$ places on a cluster of length $\ell\ge2$.
We thus arrive to the coupled linear differential equations
\beq
\frac{\dd p_\ell(t)}{\dd t}=-(\ell-1)p_\ell(t)+2\sum_{k\ge\ell+2}p_k(t)
\label{dpdt}
\eeq
for $\ell\ge1$,
with initial condition
\beq
p_\ell(0)=(1-p)^2p^\ell.
\eeq
Inserting an Ansatz of the form $p_\ell(t)=a(t)z(t)^\ell$
into~(\ref{dpdt}) yields two differential equations for $z(t)$ and $a(t)$.
We thus obtain
\beq
p_\ell(t)=(1-p\e^{-t})^2\exp(2p(\e^{-t}-1))p^\ell\e^{-(\ell-1)t},
\eeq
and so
\beqa
E(t)&=&2p\exp(2p(\e^{-t}-1))-1,
\nonumber\\
\rho(t)&=&1-p\exp(2p(\e^{-t}-1)).
\eeqa

In the limit of infinitely long times,
only inactive intervals of length $\ell=1$ survive, as should be.
Their limiting density reads
\beq
p_1(t\to\infty)=p\e^{-2p},
\eeq
and so
\beq
E_\infty=2p\e^{-2p}-1,\quad\rho_\infty=1-p\e^{-2p}
\label{erinf}
\eeq
are respectively the final energy and particle densities in the blocked configurations.

For an initially empty system $(p=1)$,
the final density (or coverage) of the dimer deposition model reads
\beq
\rho_\infty=1-\e^{-2}\approx0.864664.
\label{rhoflory}
\eeq
This is the celebrated result of Flory~\cite{flory}.

\begin{figure}[!ht]
\begin{center}
\includegraphics[angle=0,width=1\linewidth,clip=true]{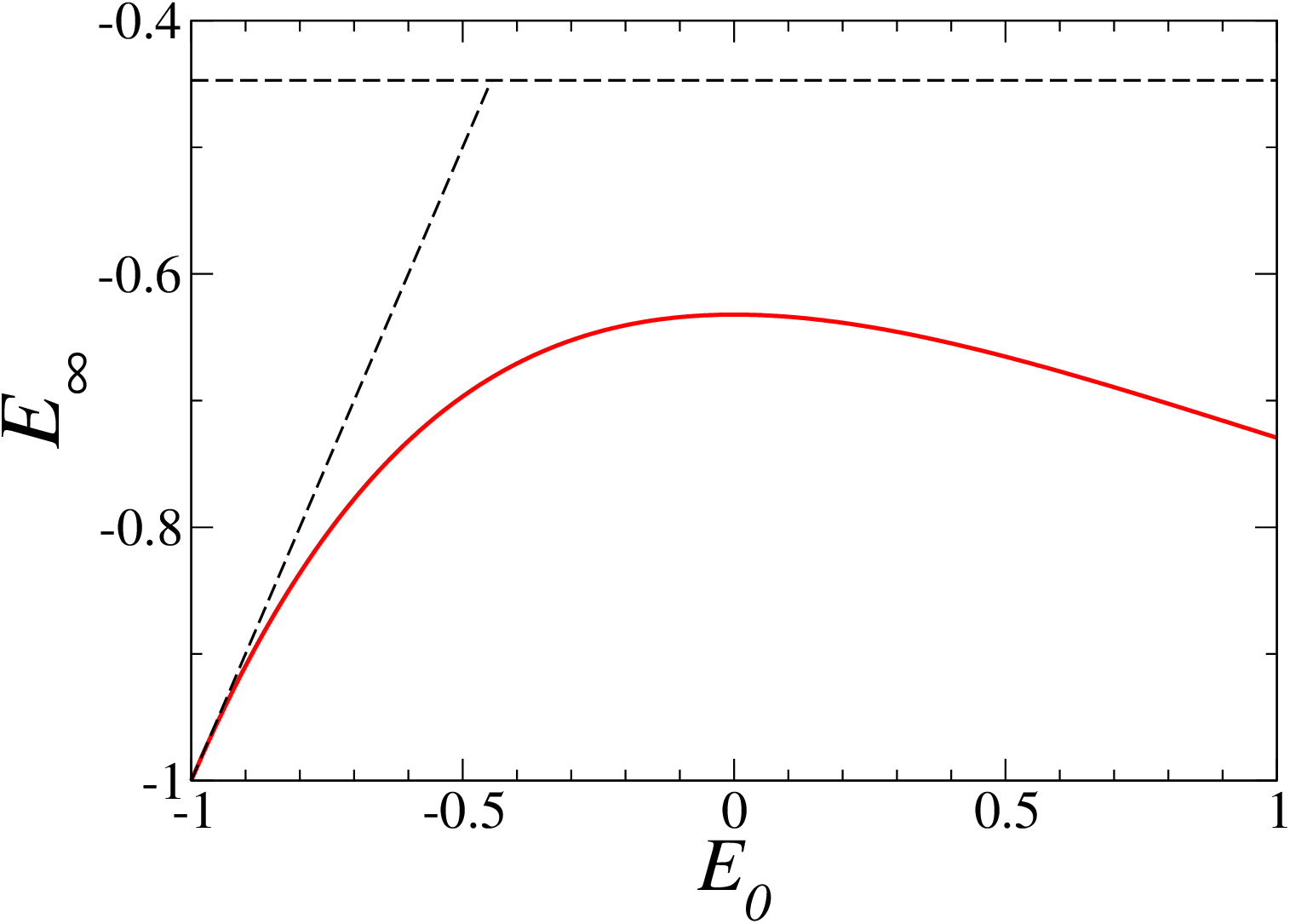}
\caption{
Final energy density $E_\infty$ of the constrained Ising chain,
as given in~(\ref{ezeires}), against initial energy density $E_0$.
Oblique dashed line: diagonal ($E_\infty=E_0$).
Horizontal dashed line: {\it a priori} value~(\ref{eap}).}
\label{dimener}
\end{center}
\end{figure}

Using~(\ref{erzero}) and~(\ref{erinf}),
the final energy density $E_\infty$ of the constrained Ising chain
can be expressed as a function of its initial energy density~$E_0$, namely
\beq
E_\infty=(E_0+1)\e^{-E_0-1}-1.
\label{ezeires}
\eeq
The final energy $E_\infty$ is plotted against $E_0$ in Figure~\ref{dimener}.
It is significantly lower than the {\it a priori} value~(\ref{eap}),
and lower than $E_0$, as should be.
Furthermore, it exhibits a non-monotonic dependence on $E_0$.
Near the ground-state energy ($E_0\to-1$),
we have $E_\infty-E_0\approx(E_0+1)^2$.
The final energy increases to the maximal value
\beq
E_\infty=\e^{-1}-1\approx-0.632120
\label{einfinf}
\eeq
for $E_0=0$, i.e., an infinite-temperature initial state,
and then decreases to the value
\beq
E_\infty=2\e^{-2}-1\approx-0.729329
\eeq
for $E_0=1$, i.e., an antiferromagnetically ordered initial state.

\subsubsection{Energy correlation function}
\label{isingdyc}

The temporal evolution of the two-point correlation introduced in~(\ref{gdef})
can also be obtained by using the shielding property.
The derivation involves the following one-cluster and two-cluster functions
\beqa
c_n(t)
&=&\prob(\underbrace{\z\cdots\z}_n),
\nonumber\\
d_{m,k,n}(t)
&=&\prob(\underbrace{\z\cdots\z}_m\,\underbrace{\cdots}_k\,\underbrace{\z\cdots\z}_n),
\eeqa
where the contents of the middle interval of length~$k$ is unspecified.
The above functions obey the coupled linear equations
\beqa
\frac{\dd c_n(t)}{\dd t}
&=&-(n-1)c_n(t)-2c_{n+1}(t),
\\
\frac{\dd d_{m,k,n}(t)}{\dd t}
&=&-(m+n-2)d_{m,k,n}(t)
\nonumber\\
&&-d_{m+1,k,n}(t)-d_{m,k,n+1}(t)
\nonumber\\
&&-d_{m+1,k-1,n}(t)-d_{m,k-1,n+1}(t),
\nonumber
\eeqa
with appropriate initial and boundary conditions.

Skipping details,
we mention the expression of the
final connected correlation $(n\ge1)$~\cite{ds}:
\beqa
C_{n,\infty}-E_\infty^2
&=&2p\e^{-2p}\frac{(-2p)^n}{n!}
\nonumber\\
&-&4p^2\e^{-2p}\sum_{m\ge n}\frac{(-2p)^m}{m!}.
\label{gcon}
\eeqa
This result exhibits the generic features of connected correlations in RSA model
already mentioned in the Introduction~\cite{evans,talbot,kinetic},
including oscillations and, more importantly, an inverse-factorial fall-off.
This peculiar kind of superexponential decay seems to have been evidenced first in~\cite{ebh}
(see~\cite{mh,hemmer} for alternative derivations).

\begin{figure}[!ht]
\begin{center}
\includegraphics[angle=0,width=1\linewidth,clip=true]{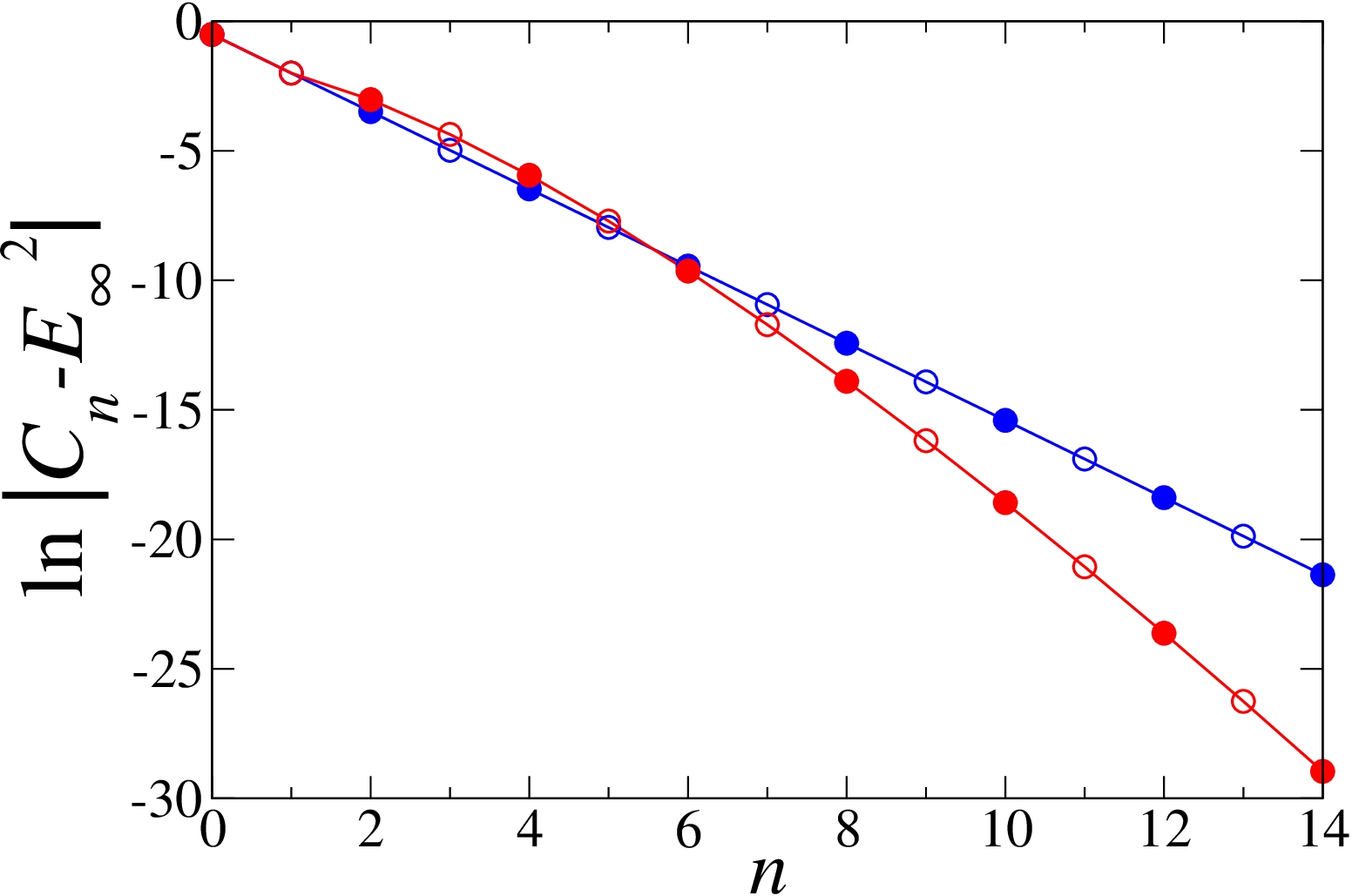}
\caption{
Comparison between the final connected correlation
$C_{n,\infty}-E_\infty^2$, given by~(\ref{gcon}) (red symbols)
and its counterpart $C_{n,\mic}-E_\infty^2$
in the {\it a priori} microcanonical ensemble at fixed energy $E$
(see~(\ref{gap})) (aligned blue symbols).
Data are plotted on a logarithmic scale against distance $n$.
They correspond to $p=1/2$ or $E_0=0$, and so $E_\infty$ is given by~(\ref{einfinf}).
Full (resp.~empty) symbols denote positive (resp.~negative) connected correlations.}
\label{dimcors}
\end{center}
\end{figure}

Figure~\ref{dimcors} shows a comparison between the final connected correlation
$C_{n,\infty}-E_\infty^2$, given by~(\ref{gcon}),
and its counterpart $C_{n,\mic}-E_\infty^2$
in the {\it a priori} microcanonical ensemble at fixed energy $E$
(see~(\ref{gap})).
In order for the comparison to be fair, the exact final energy $E_\infty$
has been used to define the {\it a priori} ensemble.
Data corresponding to an infinite-temperature initial state,
i.e., $p=1/2$ or $E_0=0$, so that $E_\infty$ is given by~(\ref{einfinf}),
are plotted on a logarithmic scale against distance~$n$.
The microcanonical ensemble correctly predicts $C_{0,\mic}=C_{0,\infty}=1$, by construction,
as well as $C_{1,\mic}=C_{1,\infty}=-2E_\infty-1$.
The two datasets cross each other between distances~5 and 6.
The superexponential decay of the exact correlations is clearly visible.

\subsubsection{Dynamical entropy}

The above method can be further extended to derive an exact expression
for the full dynamical entropy function $\Sigma_\infty(E)$,
such that the probability $P_\infty(E)$
of observing a final blocked configuration with an atypical energy density $E\ne E_\infty$
is given by a large-deviation formula of the form
\beq
P_\infty(E)\sim\exp(-N\Sigma_\infty(E)).
\eeq
The calculations involved in the derivation of dynamical large-deviation functions
in RSA models~\cite{ds,gl,kld}
are more intricate that those exposed in Sections~\ref{isingdye}
(energy density) and~\ref{isingdyc} (energy correlation).

\begin{figure}[!ht]
\begin{center}
\includegraphics[angle=0,width=1\linewidth,clip=true]{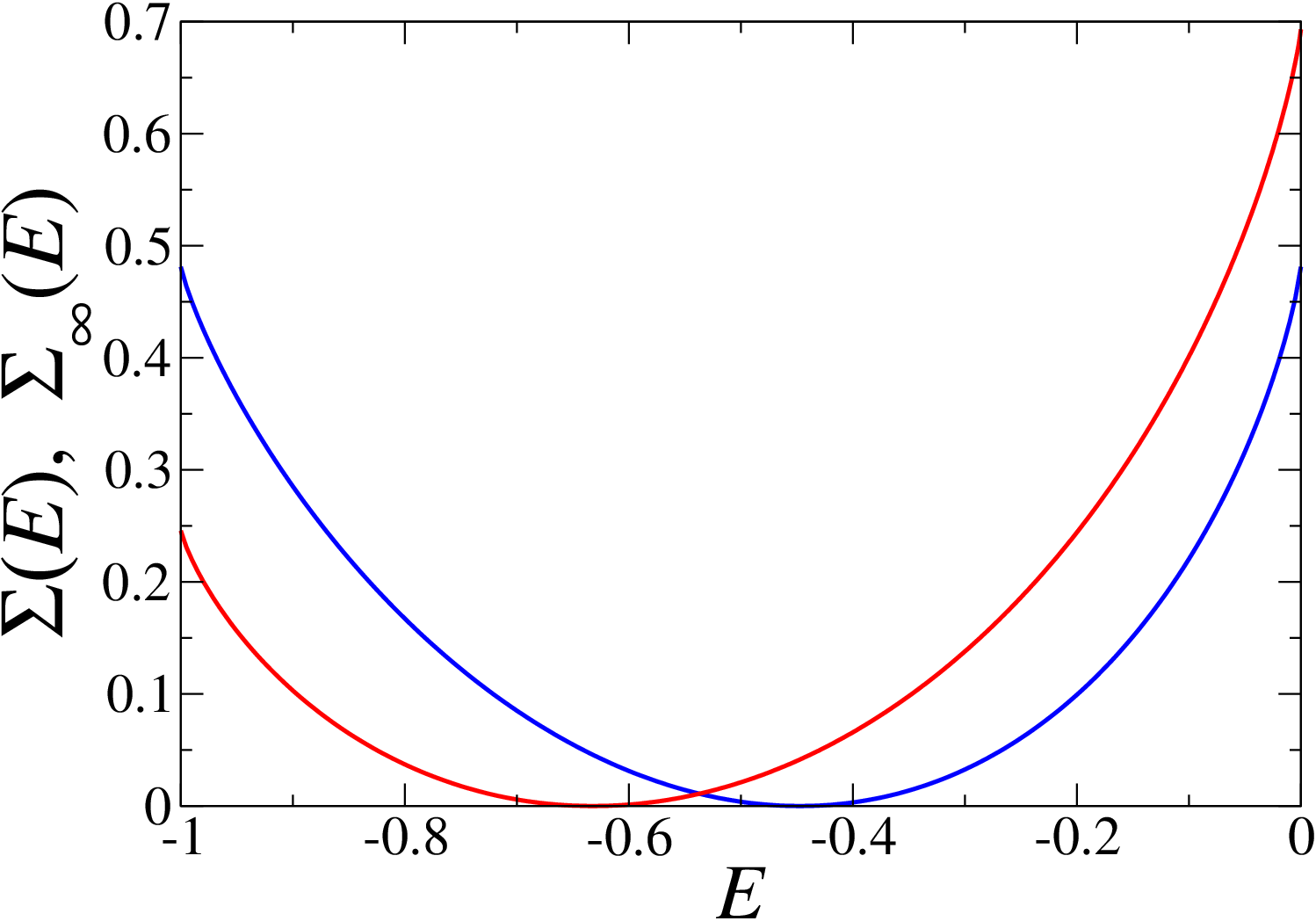}
\caption{
Red curve: dynamical entropy $\Sigma_\infty(E)$ given in~(\ref{sdy})
for $p=1/2$, corresponding to $E_0=0$, i.e., an infinite-temperature initial state.
Blue curve: {\it a priori} large-deviation function $\Sigma(E)$ given in~(\ref{sigap}).
Both quantities are plotted against the energy density $E$.}
\label{dimapdy}
\end{center}
\end{figure}

Skipping every detail,
we mention that the dynamical entropy of the constrained Ising chain
is given by the following parametric form~\cite{ds}:
\beqa
&&\Sigma_\infty=\ln z
+\frac{(1+(2p-1)z)^2-(1-z)^2\e^{4pz}}{4pz^2\e^{2pz}(2(1-p)+(2p-1)z)}
\nonumber\\
&&{\hskip 43.5pt}\times\ln\frac{1+(2p-1)z+(1-z)\e^{2pz}}{1+(2p-1)z-(1-z)\e^{2pz}},
\nonumber\\
&&E=\frac{(1+(2p-1)z)^2-(1-z)^2\e^{4pz}}
{2pz^2\e^{2pz}(2(1-p)+(2p-1)z)}-1,
\label{sdy}
\eeqa
where the parameter $z$ runs between zero and a positive maximal value $z_c$ such that
\beq
1+(2p-1)z_c+(1-z_c)\e^{2p z_c}=0.
\eeq
The dynamical entropy $\Sigma_\infty(E)$ is positive,
vanishes quadratically in the vicinity of $E=E_\infty$,
and has finite limits at both endpoints, namely
\beqa
&&\Sigma_\infty(-1)=\ln z_c,
\nonumber\\
&&\Sigma_\infty(0)=-\frac{1}{2}\ln(p(1-p)).
\eeqa
Figure~\ref{dimapdy} shows a comparison between
the dynamical entropy $\Sigma_\infty(E)$ given in~(\ref{sdy}),
for $p=1/2$, corresponding to $E_0=0$, i.e., an infinite-temperature initial state,
and the {\it a priori} large-deviation function $\Sigma(E)$, given in~(\ref{sigap}).
Both quantities are plotted against the energy density $E$.
Their endpoint values read
$\Sigma_\infty(-1)=\ln z_c\approx0.245659$,
$\Sigma_\infty(0)=\ln 2\approx0.693147$,
$\Sigma(-1)=\Sigma(0)=S_\star\approx0.481211$.

\section{The Riviera model}
\label{riv}

\subsection{Definition}
\label{rivdef}

The Riviera model is the one-dimensional variant introduced recently by Puljiz et~al.~\cite{hr3}
of a two-dimensional irreversible deposition model
introduced earlier by the same authors~\cite{hr1,hr2}.
Houses are sequentially built on plots of equal sizes along an infinitely long beach,
to be viewed as the sites of an infinite chain,
with the constraint that every house should enjoy the sunlight
from at least one of the side directions.
In other words {\it at least one} of the two neighboring plots
of each house should remain forever unbuilt.
The strand is initially empty.
New houses are introduced until a block configuration is reached.
Here is a blocked configuration on a system of length 40,
comprising 25 houses ($\u$) and 15 empty plots ($\z$):
\[
\u\u\z\u\u\z\u\u\z\u\u\z\u\u\z\u\z\u\u\z\u\z\u\z\u\u\z\z\u\u\z\u\z\u\u\z\u\u\z\u
\]

As already announced in the Introduction,
the one-dimensional Riviera model does not enjoy the shielding property.
Indeed, any newly built house imposes the constraint that
{\it at least one} of the two neighboring plots should remain forever unbuilt.
This constraint couples both semi-infinite systems on either side of the new house.
The Riviera model can therefore neither be mapped onto an RSA model,
nor (in all likelihood) be solved by analytical means.
Hereafter we first investigate
the {\it a priori} statistical ensembles of blocked configurations,
and then present two kinds of results on the statistics of attractors,
namely exact results obtained by enumeration on small systems,
and numerical results in the thermodynamic limit of very large systems.

The basic dynamical variable is the occupation~$\eta_n$ of plot number $n$:
\beq
\begin{matrix}
\u:\ \eta_n=1,\cr
\z:\ \eta_n=0.
\end{matrix}
\eeq
We shall be mostly interested in the final particle density (or coverage) of the model
when it has reached one of the many attractors which are at its disposal,
\beq
\rho_\infty=\mean{\eta_0}_\infty,
\eeq
and in the final occupation correlation
\beq
C_{n,\infty}=\mean{\eta_0\eta_n}_\infty.
\label{cdef}
\eeq

\subsection{Statistical ensembles of attractors}
\label{rivstat}

The attractors of the Riviera model
are the blocked configurations where no new house can be built.
They admit a local description in terms of two (possibly overlapping) local patterns:
\beq
\z\u\u\z,\quad\u\z\u.
\label{local}
\eeq

In line with Section~\ref{isingstat},
the statistical ensemble where all attractors are equally probable,
irrespective of their particle density~$\rho$,
is referred to as the full {\it a priori} ensemble,
whereas the ensemble consisting of all attractors with fixed density~$\rho$
is the microcanonical {\it a priori} ensemble.

\subsubsection{Transfer-matrix formalism}

Along the lines of Section~\ref{isingstat},
we introduce the Hamiltonian
\beq
\H=-\sum_n\eta_n,
\eeq
as well as the canonical ensemble
defined by introducing an effective inverse temperature~$\beta$
conjugate to $\H$.

The structure of the local patterns~(\ref{local})
leads us to introduce six partial partition functions,
defined by assigning to the occupations of the three rightmost sites
their six different permitted values.
These obey the linear recursion
\beq
\begin{pmatrix}
Z_{N+1}^{\z\z\u}\cr
Z_{N+1}^{\z\u\z}\cr
Z_{N+1}^{\z\u\u}\cr
Z_{N+1}^{\u\z\z}\cr
Z_{N+1}^{\u\z\u}\cr
Z_{N+1}^{\u\u\z}
\end{pmatrix}
=\T
\begin{pmatrix}
Z_N^{\z\z\u}\cr
Z_N^{\z\u\z}\cr
Z_N^{\z\u\u}\cr
Z_N^{\u\z\z}\cr
Z_N^{\u\z\u}\cr
Z_N^{\u\u\z}
\end{pmatrix}
\label{zrec6}
\eeq
involving the $6\times6$ transfer matrix
\beq
\T=
\begin{pmatrix}
0 & 0 & 0 & x & 0 & 0\cr
0 & 0 & 0 & 0 & 1 & 0\cr
x & 0 & 0 & 0 & x & 0\cr
0 & 0 & 0 & 0 & 0 & 1\cr
0 & x & 0 & 0 & 0 & x\cr
0 & 0 & 1 & 0 & 0 & 0
\end{pmatrix},
\label{t6}
\eeq
with the notation
\beq
x=\e^{-\beta}.
\label{xdef}
\eeq
An equivalent transfer-matrix formalism has been investigated in~\cite{hr3}.

The characteristic equation of the transfer matrix $\T$ reads
\beq
\lambda^6-x\lambda^4-x^2(\lambda^3+\lambda^2)+x^3=0.
\label{tchar}
\eeq
This equation admits the rational parametrization
\beqa
x&=&\frac{(u-1)^4(u+1)^2}{u^3},
\label{rivx}
\\
\lambda&=&\frac{(u-1)^2(u+1)}{u}.
\label{rivlam}
\eeqa
For any fixed $x$,~(\ref{rivx}) yields six values of~$u$,
and~(\ref{rivlam}) yields the corresponding six eigenvalues $\lambda_a$ of $\T$.
We denote by
$\L_a$ and $\R_a$ the left and right eigenvectors associated with~$\lambda_a$,
normalized so as to have $\L_a\cdot\R_b=\delta_{ab}$.

\begin{figure}[!ht]
\begin{center}
\includegraphics[angle=0,width=.7\linewidth,clip=true]{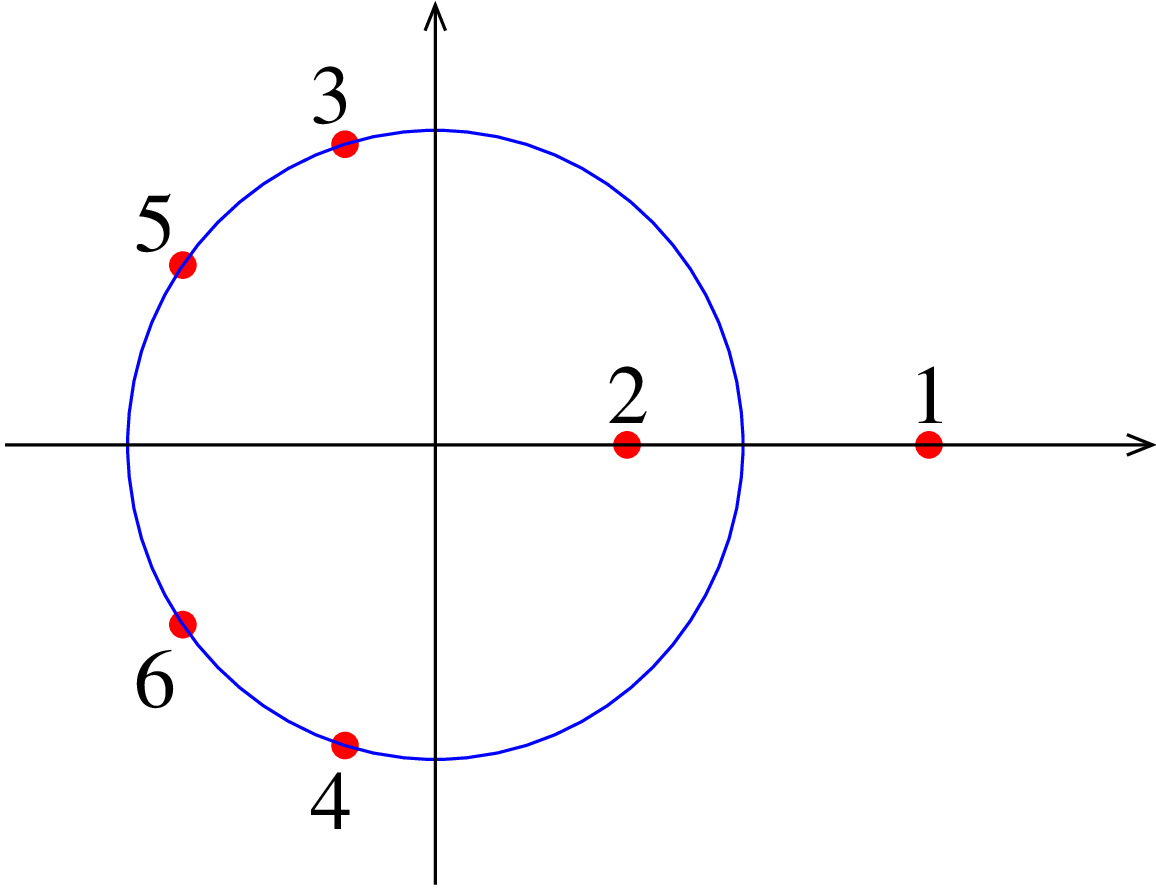}
\caption{
Structure of the spectrum of the transfer matrix $\T$ for $x<x_c$.
Red symbols: eigenvalues $\lambda_a$ labelled by $a=1,\dots,6$.
The blue circle centered at the origin with radius $\sqrt{x}$
contains the four complex eigenvalues.}
\label{spectrum}
\end{center}
\end{figure}

The parametrization~(\ref{rivx}),~(\ref{rivlam}) implies the following symmetry.
Changing $u$ into $1/u$ leaves $x$ invariant and changes $\lambda$ into $x/\lambda$.
Therefore, for any fixed $x$,
if $\lambda$ is an eigenvalue of $\T$, corresponding to the parameter~$u$,
$x/\lambda$ is another eigenvalue, corresponding to the parameter $1/u$.
The transfer matrix $\T$ has degenerate eigenvalues in two cases:
$x=0$, where the six eigenvalues collapse to the origin, and
\beq
x_c=\frac{256}{27}\approx9.481481,
\label{xcdef}
\eeq
where $\T$ has a pair of twice degenerate complex eigenvalues.
In the most relevant range $x<x_c$,
the structure of the spectrum of $\T$ is shown in Figure~\ref{spectrum}.
The largest eigenvalue $\lambda_1$
and the smallest one $\lambda_2=x/\lambda_1$ are real positive,
whereas the four other eigenvalues $\lambda_3,\dots,\lambda_6$ form two complex conjugate pairs,
with negative real parts and common modulus $\sqrt{x}$.
For $x>x_c$,
one pair of complex eigenvalues have modulus less than~$\sqrt{x}$,
whereas the other pair has modulus greater than~$\sqrt{x}$.

\subsubsection{Statistics of finite systems}

Before we investigate the configurational entropy and correlations in the thermodynamic limit,
let us focus our attention onto finite systems.

The number $\N_{N,M}$ of distinct blocked configurations
comprising $M$ houses on a finite system of $N$ sites
can be derived from the transfer-matrix formalism as follows.
The partition function
\beq
Z_N(x)=\sum_M\N_{N,M}x^M
\eeq
is given by the linear combination
corresponding to open boundary conditions at both ends
of the partial partition functions obeying the recursion~(\ref{zrec6}).
Some algebra yields the expression
\beq
Z_N(x)=\L_\open.\T^N\R_\open,
\eeq
with
\beqa
\L_\open
&=&
\begin{pmatrix}
0 & 0 & 1 & 0 & 1 & 1
\end{pmatrix},
\nonumber\\
\R_\open
&=&
\begin{pmatrix}
0 & 0 & 0 & 0 & 0 & 1
\end{pmatrix}^T,
\eeqa
where the superscript $T$ denotes transposition.
The total number of distinct blocked configurations on a system of size $N$ reads
\beq
\N_N=\sum_M\N_{N,M}=Z_N(1).
\eeq
The partition functions $Z_N(x)$ and the numbers~$\N_N$ obey the linear recursions
\beqa
Z_{N+6}(x)-xZ_{N+4}(x)&-&x^2(Z_{N+3}(x)+Z_{N+2}(x))
\nonumber\\
&+&x^3Z_N(x)=0
\label{zrec}
\eeqa
and
\beq
\N_{N+6}-\N_{N+4}-\N_{N+3}-\N_{N+2}+\N_N=0,
\eeq
as a consequence of the characteristic equa\-tion~(\ref{tchar}).

An elegant presentation of the above results in terms of their generating series
has been given in~\cite{hr3}.
With the present notations,
we have
\beqa
\Z(x,y)
&=&\sum_{M,N}\N_{N,M}x^My^N
\nonumber\\
&=&\sum_{N\ge0}Z_N(x)y^N
\nonumber\\
&=&\L_\open.(1-y\T)^{-1}\R_\open
\\
&=&\frac{1+xy+x(x-1)y^2+x^2y^3-x^3y^5}{1-xy^2-x^2(y^3+y^4)+x^3y^6}
\nonumber
\eeqa
and
\beqa
z(y)
&=&\sum_{N\ge0}\N_Ny^N
\nonumber\\
&=&\Z(1,y)
\nonumber\\
&=&\frac{1+y+y^3-y^5}{1-y^2-y^3-y^4+y^6}.
\eeqa
The above results have also been put in perspective
with several combinatorial problems in~\cite{hr3}.
In particular, the numbers $\N_N$ of blocked configurations
count a class of permutations with strongly restricted displacements.
They are listed in the OEIS under reference A080013~\cite{OEIS}
(see also~\cite{allen}).

Table~\ref{apfinite} gives the expression of the partition function $Z_N(x)$
and the total number $\N_N$ of blocked configurations for sizes $N$ up to 14.
The last line means that, on a system of $N=14$ sites,
there are altogether 91 distinct blocked configurations,
among which 50 have $M=8$ houses, 40 have $M=9$ and a single one has $M=10$,
namely $\u\u\z\u\u\z\u\u\z\u\u\z\u\u$.

\begin{table}[!ht]
\begin{center}
$$
\begin{array}{|c|c|c|}
\hline
N & Z_N(x) & \N_N\\
\hline
0 & 1 & 1\\
1 & x & 1\\
2 & x^2 & 1\\
3 & 3x^2 & 3\\
4 & x^2+2x^3 & 3\\
5 & 3x^3+x^4 & 4\\
6 & 6x^4 & 6\\
7 & 6x^4+3x^5 & 9\\
8 & x^4+10x^5+x^6 & 12\\
9 & 6x^5+10x^6 & 16\\
10 & 20x^6+4x^7 & 24\\
11 & 10x^6+22x^7+x^8 & 33\\
12 & x^6+30x^7+15x^8 & 46\\
13 & 10x^7+49x^8+5x^9 & 64\\
14 & 50x^8+40x^9+x^{10} & 91\\
\hline
\end{array}
$$
\caption{
Partition functions $Z_N(x)$ of finite systems of $N$ sites, for $N$ up to 14.
The quantity $\N_N=Z_N(1)$
is the total number of distinct blocked configurations on a system of size $N$.}
\label{apfinite}
\end{center}
\end{table}

The mean particle density $\rho_{N,\star}$
in the {\it a priori} ensemble where all blocked configurations
of the finite system of size $N$ are equally probable reads
\beq
\rho_{N,\star}=\frac{1}{NZ_N}\left(\frac{\dd Z_N}{\dd x}\right)_{x=1}.
\label{rivrhoap}
\eeq
The rational values of the {\it a priori} densities $\rho_{N,\star}$
will be given in Table~\ref{dyfinite}, for $N$ up to 14,
together with their dynamical analogues $\rho_{N,\infty}$,
defined in~(\ref{rivrhody}).

\subsubsection{Configurational entropy}

The configurational entropy $S(\rho)$ is given by~(\ref{sleg}),~(\ref{leg}),
up to the replacement of the energy density $E$ by the particle density $\rho$,
and of~$\lambda_+$ by the largest eigenvalue $\lambda_1$ of $\T$.
With the parametrization~(\ref{rivx}),~(\ref{rivlam}),
$\lambda_1$ is reached for $u$ real in the range $1<u<+\infty$.
We thus obtain the following parametric representation of the configurational entropy:
\beqa
\rho&=&\frac{2u^2+u+1}{D},
\label{rivrho}
\\
S&=&\frac{A}{D},
\label{rivs}
\\
A&=&u(3u+1)\ln u
\nonumber\\
&-&(u^2-1)\ln((u-1)^2(u+1)),
\\
D&=&3u^2+2u+3.
\label{rivd}
\eeqa

\begin{figure}[!ht]
\begin{center}
\includegraphics[angle=0,width=1\linewidth,clip=true]{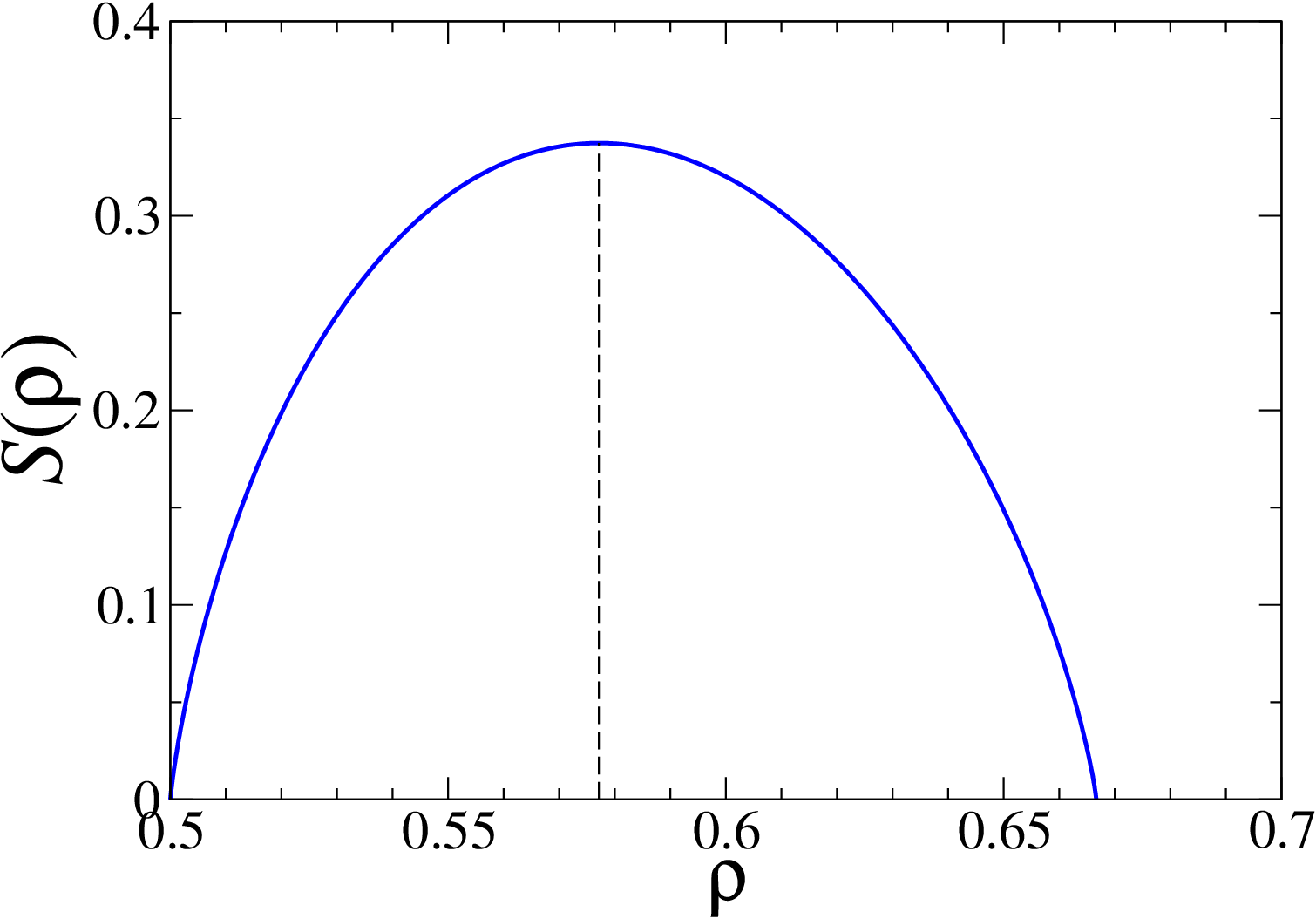}
\caption{
Configurational entropy $S(\rho)$ of the Riviera model,
as given in~(\ref{rivs}),
against particle density $\rho$ of blocked configurations.
Vertical dashed line: most probable density $\rho_\star$
yielding the maximum configurational entropy $S_\star$
(see~(\ref{rivsap}),~(\ref{rivrap})).}
\label{rivent}
\end{center}
\end{figure}

The configurational entropy $S(\rho)$ is plotted in Figure~\ref{rivent}
against the particle density $\rho$.
It vanishes at the endpoints $\rho_\min=1/2$ and $\rho_\max=2/3$,
respectively corresponding to the periodic patterns
$\u\z\u\z\u\z\cdots$ and $\u\u\z\u\u\z\u\u\z\cdots$
The maximum of the configurational entropy,
\beq
S_\star\approx0.337377,
\label{rivsap}
\eeq
is reached for $\beta=0$, i.e., $x=1$.
The corresponding values of $u$ and $\rho$ read
\beq
u_\star\approx1.963553,\quad
\rho_\star\approx0.577203.
\label{rivrap}
\eeq
The above value of $\rho_\star$ represents the most probable particle density
in the full {\it a priori} statistical ensemble.

\subsubsection{Correlation function}

The expression $C_{n,\mic}$ of the occupation correlation introduced in~(\ref{cdef})
in the microcanonical {\it a priori} ensemble at fixed density $\rho$
can be derived from the transfer-matrix formalism along the lines of Section~\ref{isingstat}.

In analogy with~(\ref{eqgap}), we have for $n\ge0$
\beq
C_{n,\mic}=\frac{B_n}{\lambda_+^n},
\label{rivcrho}
\eeq
with
\beqa
B_n
&=&\L_1\cdot\Y\T^n\Y\R_1
\label{rivb1}
\\
&=&\sum_{a=1}^6(\L_1\cdot\Y\R_a)(\L_a\cdot\Y\R_1)\lambda_a^n.
\label{rivb2}
\eeqa
Here again,
the transfer matrix~$\T$ and its eigenvalues and eigenvectors
are evaluated at the effective inverse temperature~$\beta$
related to the density~$\rho$ by~(\ref{xdef}),~(\ref{rivx}),~(\ref{rivrho}).
The diagonal matrix
\beq
\Y=
\begin{pmatrix}
1 & 0 & 0 & 0 & 0 & 0 \cr
0 & 0 & 0 & 0 & 0 & 0 \cr
0 & 0 & 1 & 0 & 0 & 0 \cr
0 & 0 & 0 & 0 & 0 & 0 \cr
0 & 0 & 0 & 0 & 1 & 0 \cr
0 & 0 & 0 & 0 & 0 & 0
\end{pmatrix}
\eeq
is the occupation operator.
We have consistently $\L_1\cdot\Y\R_1=\rho$,
so that the term with $a=1$ in~(\ref{rivb2}) yields the expected disconnected component
of the correlation, namely $\rho^2$.

The expression~(\ref{rivb1}) for the numerators $B_n$ implies that
they obey the same linear recursion as the partition functions $Z_N(x)$
(see~(\ref{zrec})), i.e.,
\beqa
B_{n+6}-xB_{n+4}&-&x^2(B_{n+3}+B_{n+2})
\nonumber\\
&+&x^3B_n=0.
\label{brec}
\eeqa
With the parametrization~(\ref{rivx}),~(\ref{rivlam}),
the recursion~(\ref{brec}) translates to a recursion relation
for the microcanonical correlations $C_{n,\mic}$ themselves:
\beqa
u^3C_{n+6,\mic}
&-&u^2C_{n+4,\mic}
\nonumber\\
&-&(u-1)^2(u+1)C_{n+3,\mic}
\nonumber\\
&-&uC_{n+2,\mic}+C_{n,\mic}=0.
\label{crec}
\eeqa

The above recursion allows one to determine recursively the correlations
at all distances from the following first six values
\beqa
C_{0,\mic}&=&\rho=\frac{2u^2+u+1}{D},
\nonumber\\
C_{1,\mic}&=&\frac{u^2}{D},
\nonumber\\
C_{2,\mic}&=&\frac{u(u+1)}{D},
\nonumber\\
C_{3,\mic}&=&\frac{2u^2-1}{D},
\nonumber\\
C_{4,\mic}&=&\frac{u^3+2u+1}{uD},
\nonumber\\
C_{5,\mic}&=&\frac{(u^2+u-1)^2}{u^2D},
\label{rivcap}
\eeqa
which can be derived from~(\ref{rivcrho}),~(\ref{rivb1}) by means of computer algebra.
The denominator $D$ has been given in~(\ref{rivd}).

The numerators of the first four correlations in~(\ref{rivcap}) are quadratic in $u$.
These quantities therefore obey two linear identities, which can be written as
\beqa
C_2&=&3\rho-2C_1-1,
\nonumber\\
C_3&=&2\rho+C_1-1.
\label{rividens}
\eeqa
All `mic' subscripts have been suppressed in the above equations for the following reason.
These identities have been derived within the microcanonical framework,
irrespective of the parameter $u$,
i.e., irrespective of the mean density $\rho$.
It will be shown below that these are constitutive identities,
which ensue from the mere structure of the blocked configurations.
In particular, the above identities hold for the final correlations $C_{n,\infty}$.

The large-distance behavior of the connected correlation
is dominated by the terms with $a=3,\dots,6$ in~(\ref{rivb2}),
involving the four complex eigenvalues of $\T$.
The phases of the latter eigenvalues are generically not rationally related to $\pi$.
The connected correlations therefore exhibit everlasting quasiperiodic oscillations,
whose hull falls off exponentially as
\beq
C_{n,\mic}-\rho^2\sim\e^{-\mu n}.
\eeq
For $x<x_c$, the structure of the spectrum shown in Figure~\ref{spectrum} implies
\beq
\mu=\ln\frac{\lambda_1}{\sqrt{x}}=\frac12\ln u.
\label{mudef}
\eeq

\subsection{Exact results on the attractors of finite systems}
\label{rivexact}

In the dynamics of the Riviera model,
sites (i.e., plots of land) are visited in a random sequential order.
The totally irreversible rules of the model ensure that a house may be built at a given site
only at the first time this site is visited,
whereas any subsequent visit is doomed to be sterile.

For an initially empty finite system of $N$ sites,
the attractor reached by a given realization of the process
therefore only depends on the order in which the various sites are visited for the first time.
It is advantageous to encode this ordering by a permutation~$\sigma$
of the $N$ site labels ($i=1,\dots,N$),
such that site $\sigma_1$ is visited first, and so on, until site $\sigma_N$ is visited last.
There are $N!$ distinct permutations, and therefore $N!$
equally probable distinct realizations of the Riviera process.
It is natural to define the dynamical weight $W(\C)$
of each attractor $\C$ as being the number of permutations
for which the system ends up in the blocked configuration $\C$.
An analogous approach based on uniform random permutations
has already been used in earlier work~\cite{theater}.

For moderate values of the system size $N$,
the~$N!$ permutations can be enumerated
and the corresponding dynamics be run by means of a computer routine,
yielding the dynamical weights of all attractors of the model,
and therefore the exact values of all observables defined in terms of those attractors.
In particular, the mean final particle density
on a system of size $N$ reads
\beq
\rho_{N,\infty}=\frac{1}{N\,N!}\sum_{\C}W(\C)M(\C),
\label{rivrhody}
\eeq
where the sum runs over all blocked configurations $\C$ of size $N$,
$W(\C)$ is the dynamical weight of $\C$ defined above,
and $M(\C)$ is the number of occupied sites (i.e., houses) in $\C$.

Table~\ref{weights} gives the dynamical weights $W(\C)$
of all attractors of the Riviera model,
as obtained by means of the above enumeration approach,
ordered by increasing values of system size $N$ and number $M$ of occupied sites,
for $N$ up to 8.
The numbers of attractors at fixed $N$ and $M$
agree with the expressions of the partition functions given in Table~\ref{apfinite}.

\begin{table}[!ht]
\begin{center}
$$
\begin{array}{|c|c|c|c|}
\hline
N & M & \C & W(\C) \\
\hline
1&1& \u & 1 \\
\hline
2&2& \u\u & 2 \\
\hline
3&2& \z\u\u & 2 \\
 & & \u\u\z & 2 \\
 & & \u\z\u & 2 \\
\hline
4&2& \z\u\u\z & 4 \\
\hline
4&3& \u\z\u\u & 10 \\
 & & \u\u\z\u & 10 \\
\hline
5&3& \z\u\u\z\u & 22 \\
 & & \u\z\u\u\z & 22 \\
 & & \u\z\u\z\u & 16 \\
\hline
5&4& \u\u\z\u\u & 60 \\
\hline
6&4& \z\u\u\z\u\u & 142 \\
 & & \u\u\z\u\u\z & 142 \\
 & & \u\z\u\z\u\u & 106 \\
 & & \u\u\z\u\z\u & 106 \\
 & & \u\z\u\u\z\u & 144 \\
 & & \u\u\z\z\u\u & 80 \\
\hline
7&4& \z\u\u\z\z\u\u & 280 \\
 & & \u\u\z\z\u\u\z & 280 \\
 & & \z\u\u\z\u\z\u & 318 \\
 & & \u\z\u\z\u\u\z & 318 \\
 & & \z\u\u\z\u\u\z & 364 \\
 & & \u\z\u\z\u\z\u & 272 \\
\hline
7&5& \u\z\u\u\z\u\u & 1158 \\
 & & \u\u\z\u\u\z\u & 1158 \\
 & & \u\u\z\u\z\u\u & 892 \\
\hline
8&4& \z\u\u\z\z\u\u\z & 1120 \\
\hline
8&5& \z\u\u\z\u\z\u\u & 3114\\
 & & \u\u\z\u\z\u\u\z & 3114\\
 & & \z\u\u\z\u\u\z\u & 3496\\
 & & \u\z\u\u\z\u\u\z & 3496\\
 & & \u\z\u\z\u\z\u\u & 2386\\
 & & \u\u\z\u\z\u\z\u & 2386\\
 & & \u\z\u\z\u\u\z\u & 2768\\
 & & \u\z\u\u\z\u\z\u & 2768\\
 & & \u\z\u\u\z\z\u\u & 2464\\
 & & \u\u\z\z\u\u\z\u & 2464\\
\hline
8&6& \u\u\z\u\u\z\u\u & 10744\\
\hline
\end{array}
$$
\caption{
Dynamical weights $W(\C)$ of all attractors of the Riviera model,
ordered by increasing values of system size $N$ and number $M$ of occupied sites,
for $N$ up to 8.}
\label{weights}
\end{center}
\end{table}

\begin{figure}[!ht]
\begin{center}
\includegraphics[angle=0,width=1\linewidth,clip=true]{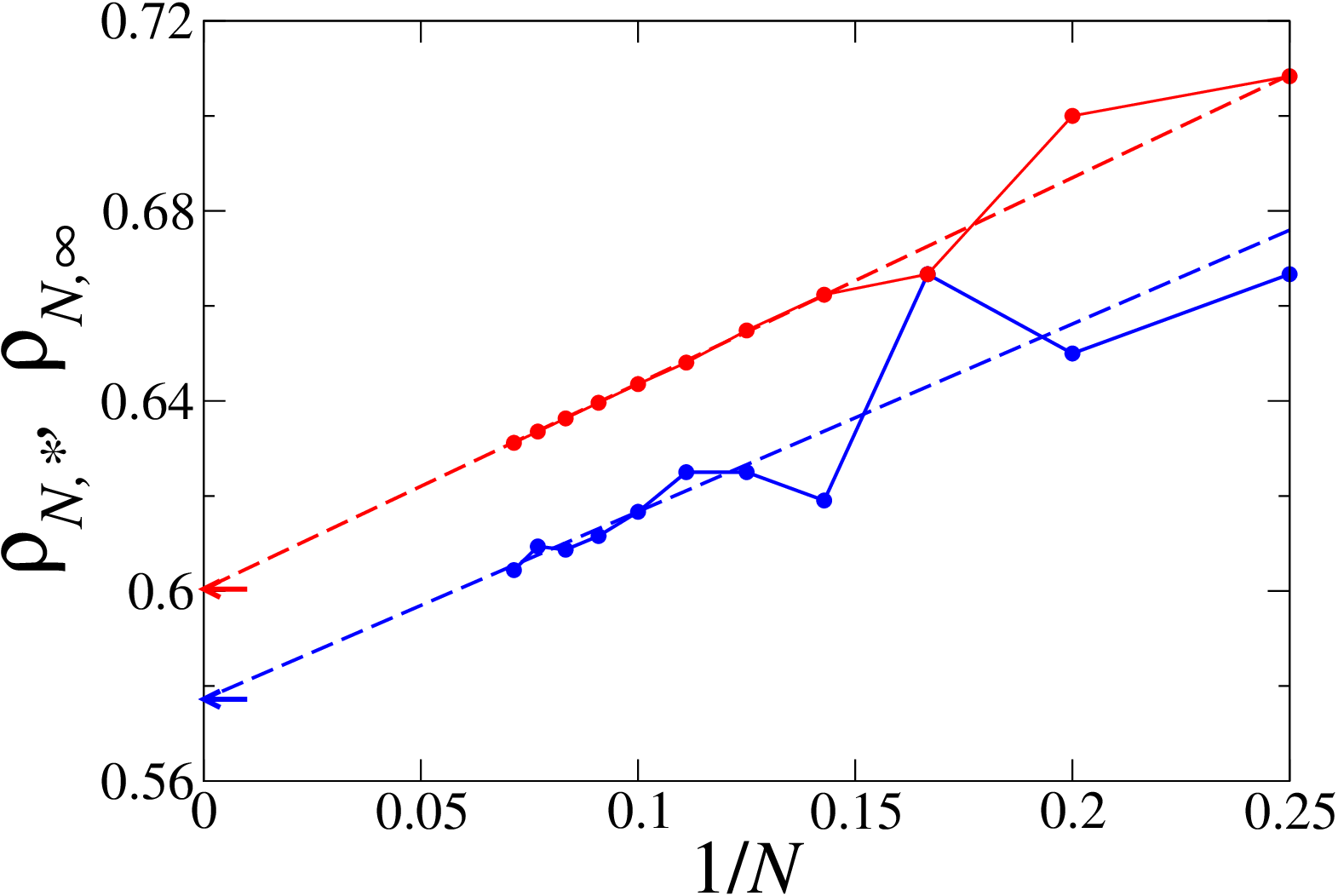}
\caption{
Blue symbols: {\it a priori} densities $\rho_{N,\star}$ defined in~(\ref{rivrhoap}).
Red symbols: final densities $\rho_{N,\infty}$ defined in~(\ref{rivrhody}).
Data are taken from Table~\ref{dyfinite} and plotted against $1/N$,
for sizes $N$ ranging from 4 to 14.
Arrows: respective limits $\rho_\star$ given in~(\ref{rivrap})
and $\rho_\infty$ given in Table~\ref{nums}.
Dashed lines: guides to the eye demonstrating the asymptotic $1/N$ convergence
of both datasets.}
\label{rhocv}
\end{center}
\end{figure}

The exact rational values of the {\it a priori} densities $\rho_{N,\star}$,
defined in~(\ref{rivrhoap}) in terms of the partition functions $Z_N$,
and of the final densities $\rho_{N,\infty}$,
defined in~(\ref{rivrhody}) in terms of the dynamical weights $W(\C)$,
are given in Table~\ref{dyfinite} and plotted against $1/N$ in Figure~\ref{rhocv}
for system sizes $N$ up to 14,
the largest system size for which the systematic enumeration approach has been carried out.

\begin{table}[!ht]
\begin{center}
$$
\begin{array}{|c|c|c|}
\hline
N & \rho_{N,\star} & \rho_{N,\infty}\\
\hline
1 & 1 & 1\\
2 & 1 & 1\\
3 & 2/3 & 2/3\\
4 & 2/3 & 17/24\\
5 & 13/20 & 7/10\\
6 & 2/3 & 2/3\\
7 & 13/21 & 2921/4410\\
8 & 5/8 & 8801/13440\\
9 & 5/8 & 7559/11664\\
10 & 37/60 & 38921/60480\\
11 & 74/121 & 2340323/3659040\\
12 & 14/23 & 152397907/239500800\\
13 & 39/64 & 6410910971/10118908800\\
14 & 55/91 & 61139821/96864768\\
\hline
\end{array}
$$
\caption{
Exact rational values of the {\it a priori} densities $\rho_{N,\star}$
defined in~(\ref{rivrhoap}),
and of the final densities $\rho_{N,\infty}$
defined in~(\ref{rivrhody}), for system sizes $N$ up to 14.}
\label{dyfinite}
\end{center}
\end{table}

The complexity of the expressions of the final densities
increases much faster than
that of the {\it a priori} ones.
This testifies the non-triviality of the dynamics of the Riviera model.
Except for the system sizes $N=1$, 2, 3 and 6,
where all attractors have the same numbers of houses,
respectively $M=1$, 2, 2 and 4,
the final density $\rho_{N,\infty}$ is larger than the {\it a priori} one by a few percent.
This means that the dynamics of the Riviera model is able to reach on average
slightly more compact configurations than a flat ensemble.
We shall come back to this point in the Discussion.

For very large system sizes,
the {\it a priori} densities $\rho_{N,\star}$ converge to the exactly known limit $\rho_\star$
given in~(\ref{rivrap}),
whereas the final densities $\rho_{N,\infty}$ converge to the limit $\rho_\infty$,
given in Table~\ref{nums},
which is only known through extensive numerical simulations.
Figure~\ref{rhocv} demonstrates that both datasets
exhibit some irregular behavior for the smaller sizes,
and smoothly converge to their respective limits as $1/N$.

\subsection{Numerical results in the thermodynamic limit}
\label{rivnum}

We end our investigation of the attractors of the Riviera model
by describing the outcomes of extensive numerical simulations
on systems of sizes $N=100$ and 200, comprising a total of $3.10^{12}$ sites.
We have measured the final occupation correlation $C_{n,\infty}$
for distances $n$ up to 13,
beyond which the connected correlation is too small to be measured accurately.

Before we present numerical results,
we wish to emphasize that the mean particle density and nearest-neighbor occupation correlation,
i.e.,
\beq
C_{0,\infty}=\rho_\infty,\quad
C_{1,\infty},
\label{basic}
\eeq
can serve as a basis to express the main local observables characterizing
the blocked configurations.
First of all, the two local densities and the four nearest-neighbor correlations read
\beqa
\mean{\u}_\infty
&=&\rho_\infty,
\nonumber\\
\mean{\z}_\infty
&=&1-\rho_\infty,
\nonumber\\
\mean{\u\u}_\infty
&=&C_{1,\infty},
\nonumber\\
\mean{\u\z}_\infty
&=&\mean{\z\u}_\infty=\rho_\infty-C_{1,\infty},
\nonumber\\
\mean{\z\z}_\infty
&=&C_{1,\infty}-2\rho_\infty+1.
\label{num1}
\eeqa

Using the above expressions,
we can prove that the identities~(\ref{rividens}) are constitutive to the model,
by using only the local structure of the blocked configurations,
as given by the patterns~(\ref{local}), as well as their translational invariance.
We have indeed
\beqa
C_{2,\infty}&=&\mean{\u\z\u}_\infty
\nonumber\\
&=&\mean{\u\z}_\infty-\mean{\u\z\z}_\infty
\nonumber\\
&=&\mean{\u\z}_\infty-\mean{\z\z}_\infty
\nonumber\\
&=&3\rho_\infty-2C_{1,\infty}-1,
\nonumber\\
C_{3,\infty}&=&\mean{\u\z\z\u}_\infty+\mean{\u\u\z\u}_\infty+\mean{\u\z\u\u}_\infty
\nonumber\\
&=&\mean{\z\z}_\infty+2\mean{\u\u\z\u}_\infty
\nonumber\\
&=&\mean{\z\z}_\infty+2(\mean{\u\z\u}_\infty-\mean{\z\u\z}_\infty)
\nonumber\\
&=&2\rho_\infty+C_{1,\infty}-1.
\eeqa

The densities $\rho_\infty(\u)$ and $\rho_\infty(\u\u)$
of clusters of one and two consecutive occupied sites (houses)
and the densities $\rho_\infty(\z)$ and $\rho_\infty(\z\z)$
of clusters of one and two consecutive empty sites read
\beqa
\rho_\infty(\u)&=&\rho_\infty-2C_{1,\infty},
\nonumber\\
\rho_\infty(\u\u)&=&C_{1,\infty},
\nonumber\\
\rho_\infty(\z)&=&3\rho_\infty-2C_{1,\infty}-1,
\nonumber\\
\rho_\infty(\z\z)&=&C_{1,\infty}-2\rho_\infty+1.
\label{num2}
\eeqa
The total density of clusters of occupied (or empty) sites therefore reads
\beqa
\rho_\infty(\cluster)
&=&\rho_\infty(\u)+\rho_\infty(\u\u)
\nonumber\\
&=&\rho_\infty(\z)+\rho_\infty(\z\z)
\nonumber\\
&=&\mean{\u\z}_\infty
=\mean{\z\u}_\infty
\nonumber\\
&=&\rho_\infty-C_{1,\infty}.
\label{num3}
\eeqa

Table~\ref{nums} gives the numerical values of the two basic observables
introduced in~(\ref{basic}),
as well as those of the local observables
given in~(\ref{num1}),~(\ref{num2}),~(\ref{num3}).
In these and subsequent numerical values,
statistical errors are estimated to be of the order of $10^{-6}$,
i.e., one unit in the last significant digit.

\begin{table}[!ht]
\begin{center}
$$
\begin{array}{|c|c|}
\hline
\rho_\infty & 0.600385 \\
C_{1,\infty} & 0.237565 \\
\hline
\mean{\u}_\infty & 0.600385 \\
\mean{\z}_\infty & 0.399615 \\
\hline
\mean{\u\u}_\infty & 0.237565 \\
\mean{\u\z}_\infty & 0.362820 \\
\mean{\z\z}_\infty & 0.036795 \\
\hline
\rho_\infty(\u) & 0.125255 \\
\rho_\infty(\u\u) & 0.237565 \\
\rho_\infty(\z) & 0.326025 \\
\rho_\infty(\z\z) & 0.036795 \\
\rho_\infty(\cluster) & 0.362820\\
\hline
\end{array}
$$
\caption{
Numerical values of the two basic obser\-vables introduced in~(\ref{basic})
and of all local observables given in~(\ref{num1}),~(\ref{num2}),~(\ref{num3}).}
\label{nums}
\end{center}
\end{table}

Let us turn to the comparison of numerical data
to the predictions of the {\it a priori} ensembles.
The final density $\rho_\infty\approx0.600385$ (see~Table~\ref{nums})
is larger than the most probable density $\rho_\star\approx0.577203$
in the full {\it a priori} ensemble (see~(\ref{rivrap}))
by a small but significant amount,
\beq
\rho_\infty-\rho_\star\approx0.023182.
\label{drho}
\eeq
It is interesting to observe that $\rho_\star$ is slightly below
the middle of the range $[1/2,2/3]$ of permitted densities,
namely $\rho_\middle=7/12\approx0.583333$,
whereas $\rho_\infty$ is slightly larger than $\rho_\middle$.
These inequalities are rather general.
We shall come back to this point in the Discussion
(see~(\ref{rhoins}) and Table~\ref{comparison}).

The microcanonical {\it a priori} ensemble of blocked configurations
whose mean density equals the final density $\rho_\infty$
corresponds to the following values
of the parameters $u$ and $x$ (see~(\ref{rivx}),~(\ref{rivrho}),~(\ref{rivd})):
\beq
u_\infty\approx2.574600,\quad
x_\infty\approx4.602640.
\label{rivux}
\eeq
The small density difference~(\ref{drho})
translates into a sizeable difference between $x_\infty$
and the value of $x$ maximizing the configurational entropy,
i.e., $x_\star=1$ by construction.
We nevertheless have $x_\infty<x_c$ (see~(\ref{xcdef})).
The correlation $C_{n,\mic}$ in the microcanonical ensemble at density $\rho_\infty$
can be readily obtained for all distances $n$ by inserting
the value of $u_\infty$ given in~(\ref{rivux})
into the initial values~(\ref{rivcap}) and the recursion~(\ref{crec}).

Table~\ref{cors} presents a comparison between the numerical values of the final correlations
$C_{n,\infty}$ and the exact microcanonical ones $C_{n,\mic}$ for distances $n=1$, 2 and 3,
where correlations obey the identities~(\ref{rividens}).
The differences $C_{n,\infty}-C_{n,\mic}$, given in the last column,
are very small.
More importantly, they are found to be proportional to $+1$, $-2$ and $+1$,
in agreement with the identities~(\ref{rividens}).
This parameter-free agreement to all significant digits
provides a quantitative check of the accuracy of our numerical results.

\begin{table}[!ht]
\begin{center}
$$
\begin{array}{|c|c|c|c|}
\hline
n & C_{n,\infty} & C_{n,\mic} & C_{n,\infty}-C_{n,\mic} \\
\hline
1 & 0.237565 & 0.236440 & +0.001125 \\
2 & 0.326025 & 0.328275 & -0.002250 \\
3 & 0.438335 & 0.437210 & +0.001125 \\
\hline
\end{array}
$$
\caption{
Numerical values of the final correlations $C_{n,\infty}$,
exact microcanonical correlations $C_{n,\mic}$,
and differences $C_{n,\infty}-C_{n,\mic}$,
for distances $n=1$, 2 and 3,
where correlations obey the identities~(\ref{rividens}).}
\label{cors}
\end{center}
\end{table}

\begin{figure}[!ht]
\begin{center}
\includegraphics[angle=0,width=1\linewidth,clip=true]{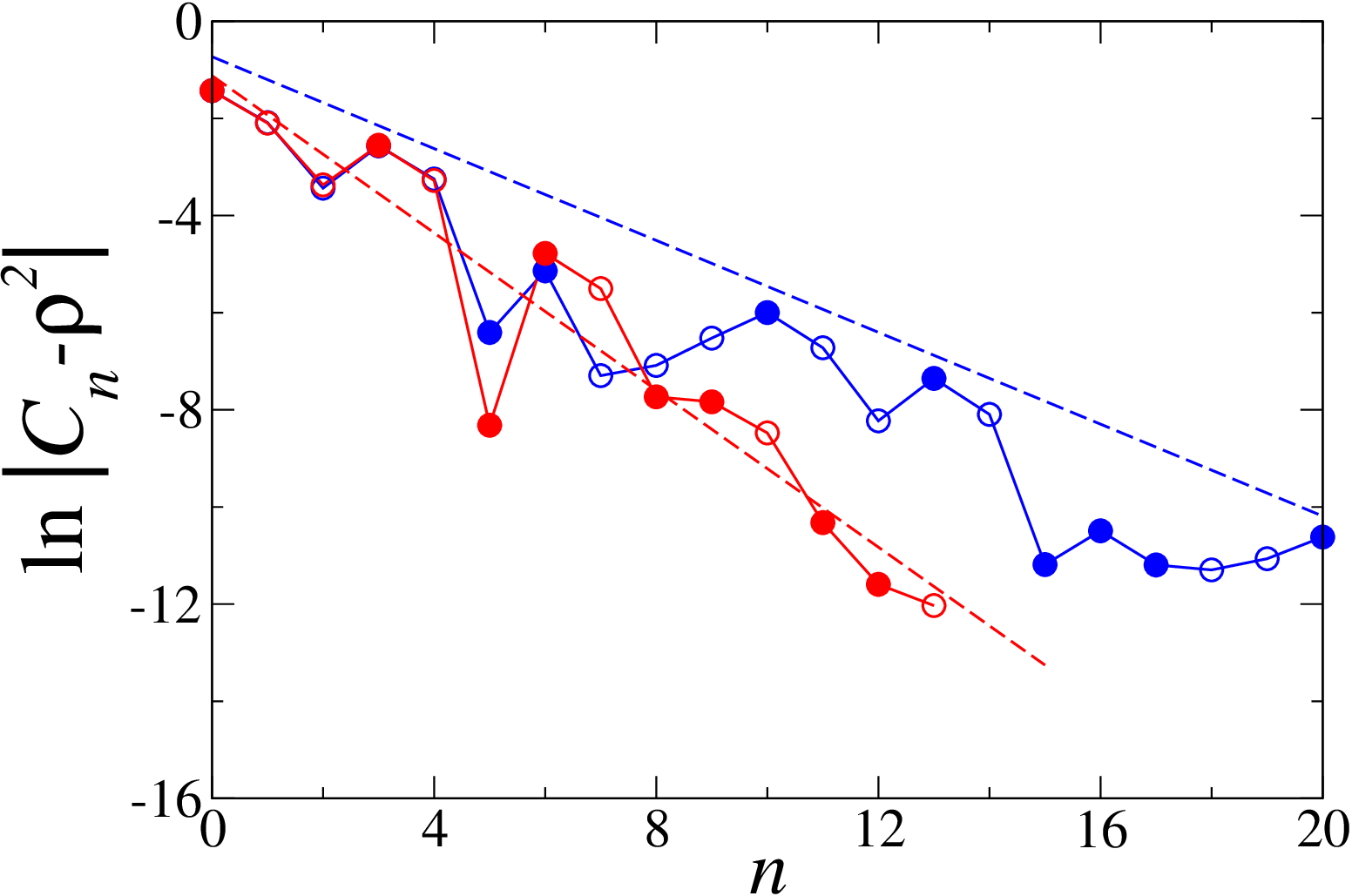}
\caption{
Comparison between the final connected correlation
$C_{n,\infty}-\rho_\infty^2$ (numerical values, red symbols)
and its counterpart $C_{n,\mic}-\rho_\infty^2$
in the microcanonical {\it a priori} ensemble at fixed density $\rho_\infty$
(exact values, blue symbols).
Data are plotted on a logarithmic scale against distance $n$.
Full (resp.~empty) symbols denote positive (resp.~negative) connected correlations.
The blue dashed line has the exact slope~(\ref{mumic}).
The red dashed with slope 0.81 is the outcome of a least-square~fit.}
\label{rivcors}
\end{center}
\end{figure}

Figure~\ref{rivcors} shows a comparison between the final connected correlation
$C_{n,\infty}-\rho_\infty^2$ (numerical values, red symbols)
and its counterpart $C_{n,\mic}-\rho_\infty^2$
in the microcanonical ensemble at density $\rho_\infty$
(exact values, blue symbols).
Data are plotted on a logarithmic scale against distance $n$.
Full (resp.~empty) symbols denote positive (resp.~negative) connected correlations.
The microcanonical connected correlations exhibit everlasting irregular oscillations
on this logarithmic scale,
whose hull falls off exponentially with decay rate
(inverse correlation length)
\beq
\mu_\mic=\frac{1}{2}\ln u_\infty\approx0.472847
\label{mumic}
\eeq
(blue dashed line), as predicted in~(\ref{mudef}).
The final connected correlations exhibit smaller oscillations,
except for an outlier at distance 5.
These oscillations seem to be damped on this logarithmic scale,
in the sense that their amplitude exhibits a slow decay,
at variance with the microcanonical ones.
Their hull is observed to fall off roughly twice faster, with an apparent decay rate
\beq
\mu_\infty\approx0.81
\label{mudy}
\eeq
(red dashed line).
The observed period three in the signs of final connected correlations
is probably a transient phenomenon with no intrinsic meaning.

\begin{table*}[!t]
\begin{center}
$$
\begin{array}{|c|c|c|c|c|c|c|}
\hline
\hbox{Number} & \hbox{Model} & \rho_\min & \rho_\max & \rho_\middle & \rho_\star & \rho_\infty \\
\hline
1 & \hbox{Riviera} & 0.5 & 0.666666 & 0.583333 & 0.577203 & 0.600385 \\
2 & \hbox{Dimers} & 0.666666 & 1 & 0.833333 & 0.822991 & 0.864664 \\
3 & \hbox{Ising-Glauber} & 0.5 & 1 & 0.75 & 0.723607 & 0.816060 \\
4 & \hbox{Ising-Kawasaki} & 0.333333 & 1 & 0.666666 & 0.618420 & {\it 0.637043} \\
5 & \hbox{Paramagnetic Ising} & 0.5 & 1 & 0.75 & 0.723607 & {\it 0.696734} \\
\hline
\end{array}
$$
\caption{
Various characteristic densities for the Riviera model
and four kinetically constrained Ising chains whose dynamics can be mapped onto an RSA model:
$\rho_\min$ and $\rho_\max$ are the endpoints of the range of permitted densities,
$\rho_\middle$ is the midpoint of the latter range,
$\rho_\star$ is the most probable density in the full {\it a priori} statistical ensemble,
$\rho_\infty$ is the final density of the attractors of the irreversible dynamics
launched from a prescribed initial state,
be it either empty or corresponding to infinite temperature (see text for details).
All data obey the inequalities~(\ref{rhoins}),
except the last two entries of the rightmost column (slanted numbers).}
\label{comparison}
\end{center}
\end{table*}

\section{Discussion}
\label{disc}

The goal of the present paper was twofold.
We have given a concise review of the theory of
one-dimensional spin models with kinetic constraints,
exemplified by the case of the constrained Ising chain,
and put this body of knowledge
in perspective with a range of novel results on the Riviera model
introduced recently by Puljiz et~al.~\cite{hr3}.

This work has evidenced
that there are both significant analogies and differences
between the one-dimensional kinetically constrained models
whose fully irreversible zero-temperature dynamics can be mapped onto an RSA model,
and the Riviera model,
whose fully irreversible dynamics
does not enjoy the characteristic shielding property of those RSA models.
It is worth noticing that the bilateral avatar of the Riviera model,
defined by the requirement that {\it both} neighboring plots
of each house should remain forever unbuilt,
amounts to a deposition model with excluded volume described in~\cite{evans}.
The latter model is equivalent to the dimer deposition model on the dual lattice.
It therefore belongs to the above class of kinetically constrained models.

Among the common features shared by the Riviera model and the class of RSA models,
we wish to underline that the final particle density~$\rho_\infty$,
i.e., the particle density of the attractors reached by the dynamics,
is slightly different from the most probable density $\rho_\star$
in the full {\it a~priori} ensemble of blocked configurations.
One can therefore speak of a universally weak violation of the Edwards flatness hypothesis.
In most cases we have
\beq
\rho_\star<\rho_\middle<\rho_\infty,
\label{rhoins}
\eeq
where $\rho_\middle=(\rho_\min+\rho_\max)/2$
is the midpoint of the range of permitted densities.
The above rule of thumb expresses that the statistical
ensemble tends to favor lower-density blocked configurations,
whereas the dynamics tends to favor higher-density blocked configurations.
This is illustrated by Table~\ref{comparison},
giving the values of $\rho_\min$, $\rho_\max$, $\rho_\middle$,
$\rho_\star$ and $\rho_\infty$
for the Riviera model and four characteristic one-dimensional kinetic spin
models whose zero-temperature dynamics can be mapped onto an RSA model.
The detailed contents of Table~\ref{comparison} are as follows.
Model~1 is the Riviera model, investigated extensively in Section~\ref{riv}.
The exact value of $\rho_\star$ and the numerical value of~$\rho_\infty$
are respectively given in~(\ref{rivrap}) and in Table~\ref{nums}.
For the four subsequent models,
both $\rho_\star$ and $\rho_\infty$ are known exactly.
Model~2 is the problem of dimer deposition.
The value of $\rho_\star$ is not related to~(\ref{eap}),
because here the system is assumed to be initially empty~\cite{kld}.
The value of $\rho_\infty$ is the celebrated result of Flory (see~(\ref{rhoflory})).
The last three models concern the Ising chain
with an infinite-temperature initial state.
Model~3 is the ferromagnetic Ising chain with kinetically constrained
zero-temperature Glauber dynamics,
investigated extensively in Section~\ref{ising}
(see~(\ref{eap}) and~(\ref{einfinf}), with $E=1-2\rho$).
Model~4 is the ferromagnetic Ising chain with kinetically constrained
zero-temperature Kawasaki dynamics~\cite{pf,ef,kpri,klin,kkra},
whereas model~5 is the paramagnetic Ising chain in a uniform magnetic field
subjected to zero-temperature either symmetrically
or asymmetrically constrained dynamics~\cite{pfa,pje,pse}.
In the last two cases, the final density $\rho_\infty$ (slanted numbers)
is smaller than $\rho_\middle$ and therefore does not obey the inequality~(\ref{rhoins}).

The most salient qualitative difference between the Riviera model
and the kinetically constrained models in the RSA class
concerns their final correlation functions.
This unlikeness was somewhat to be expected,
since the Riviera model does not enjoy the shielding property.
The numerical data shown in Figure~\ref{rivcors}
strongly suggest that the connected correlation function
$C_{n,\infty}-\rho_\infty^2$ falls off exponentially,
with a decay rate $\mu\approx0.81$ (see~(\ref{mudy})).
This observed exponential decay is in stark contrast with the superexponential,
inverse-factorial decay of connected correlations
that is universally met in one-dimensional RSA models.
Exponentially decaying correlations are rather viewed
as a characteristic of either thermal or otherwise equilibrated systems.
From this viewpoint the Riviera model thus appears as a chimera with,
on the one hand, a fully irreversible dynamics leading to metastability
with an exponentially large number of attractors and,
on the other hand, exponentially decaying correlations germane to those
observed at thermal equilibrium.

\subsubsection*{Acknowledgments}

\small
It is a pleasure to dedicate this paper to our colleague and friend Malte Henkel
on the occasion of his 60th birthday.
We acknowledge with thanks useful exchanges with
Mate Puljiz, Stjepan \v{S}ebek and Josip \v{Z}ubrini\'c.

\subsubsection*{Author contribution statement}

\small
Both authors contributed equally to the present work,
were equally involved in the preparation of the manuscript,
and have read and approved the final manuscript.

\subsubsection*{Data availability statement}

\small
Data sharing not applicable to this article as no datasets were generated or analysed
during the current study.

\bibliography{paper.bib}

\end{document}